\documentstyle[spie,epsf]{article} 
\input{psfig}   

\title{
Design Considerations for Large Detector Arrays\\
on Submillimeter-wave Telescopes
}

\author{Antony A. Stark
\skiplinehalf 
Smithsonian Astrophysical Observatory, 60 Garden St. MS78, Cambridge, MA 02138, USA
}

\authorinfo{Other author information: 
E-mail: {\tt aas@cfa.harvard.edu}; WWW: {\tt http://cfa-www.harvard.edu/$\sim$aas/tenmeter}}

\begin{document} 
  \maketitle 

\def\aj{{AJ}}			
\def\araa{{Ann. Rev. Astron.\& Astrophys.}}		
\def\apj{{ApJ}}			
\def\apjl{{ApJ (Letters)}}		
\def\apjs{{ApJS}}		
\def\ao{{Appl.Optics}}		
\def\apss{{Ap\&SS}}		
\def\aap{{A\&A}}		
\def\aapr{{A\&A~Rev.}}		
\def\aaps{{A\&AS}}		
\def\azh{{AZh}}			
\def\baas{{BAAS}}		
\def\jrasc{{JRASC}}		
\def\memras{{MmRAS}}		
\def\mnras{{MNRAS}}		
\def\pra{{Phys.Rev.A}}		
\def\prb{{Phys.Rev.B}}		
\def\prc{{Phys.Rev.C}}		
\def\prd{{Phys.Rev.D}}		
\def\prl{{Phys.Rev.Lett}}	
\def\pasp{{PASP}}		
\def\pasj{{PASJ}}		
\def\qjras{{QJRAS}}		
\def\skytel{{S\&T}}		
\def\solphys{{Solar~Phys.}}	
\def\sovast{{Soviet~Ast.}}	
\def\ssr{{Space~Sci.Rev.}}	
\def\zap{{ZAp}}			
\let\astap=\aap
\let\apjlett=\apjl
\let\apjsupp=\apjs
\def\deg{\hbox{$^\circ$}}
\def\sun{\hbox{$\odot$}}
\def\earth{\hbox{$\oplus$}}
\def\la{\mathrel{\hbox{\rlap{\hbox{\lower4pt\hbox{$\sim$}}}\hbox{$<$}}}}
\def\ga{\mathrel{\hbox{\rlap{\hbox{\lower4pt\hbox{$\sim$}}}\hbox{$>$}}}}
\def\sq{\hbox{\rlap{$\sqcap$}$\sqcup$}}
\def\arcmin{\hbox{$^\prime$}}
\def\arcsec{\hbox{$^{\prime\prime}$}}
\def\fd{\hbox{$.\!\!^{\rm d}$}}
\def\fh{\hbox{$.\!\!^{\rm h}$}}
\def\fm{\hbox{$.\!\!^{\rm m}$}}
\def\fs{\hbox{$.\!\!^{\rm s}$}}
\def\fdg{\hbox{$.\!\!^\circ$}}
\def\farcm{\hbox{$.\mkern-4mu^\prime$}}
\def\farcs{\hbox{$.\!\!^{\prime\prime}$}}
\def\fp{\hbox{$.\!\!^{\scriptscriptstyle\rm p}$}}
\def\micron{\hbox{$\mu$m}}
\def\onehalf{\hbox{$\,^1\!/_2$}}	
\def\onethird{\hbox{$\,^1\!/_3$}}
\def\twothirds{\hbox{$\,^2\!/_3$}}
\def\onequarter{\hbox{$\,^1\!/_4$}}
\def\threequarters{\hbox{$\,^3\!/_4$}}
\def\ubvr{\hbox{$U\!BV\!R$}}		
\def\ub{\hbox{$U\!-\!B$}}		
\def\bv{\hbox{$B\!-\!V$}}		
\def\vr{\hbox{$V\!-\!R$}}		
\def\ur{\hbox{$U\!-\!R$}}		
\def\gs{\mathrel{\raise0.35ex\hbox{$\scriptstyle >$}\kern-0.6em 
\lower0.40ex\hbox{{$\scriptstyle \sim$}}}}
\def\ls{\mathrel{\raise0.35ex\hbox{$\scriptstyle <$}\kern-0.6em 
\lower0.40ex\hbox{{$\scriptstyle \sim$}}}}
\def\ci{C~{I}}
\def\cii{C~{II}}
\def\cplus{\alwaysmath{{\rm C^+}}}
\newcount\lecurrentfam
\def\LaTeX{\lecurrentfam=\the\fam \leavevmode L\raise.42ex
\hbox{$\fam\lecurrentfam\scriptstyle\kern-.3em A$}\kern-.15em\TeX}
\def\plotone#1{\centering \leavevmode
\epsfxsize=\textwidth \epsfbox{#1}}
\def\plottwo#1#2{\centering \leavevmode
\epsfxsize=.45\textwidth \epsfbox{#1} \hfil
\epsfxsize=.45\textwidth \epsfbox{#2}}
\def\plotfiddle#1#2#3#4#5#6#7{\centering \leavevmode
\vbox to#2{\rule{0pt}{#2}}
\includegraphics{#1}}
\begin{abstract}
The emerging technology of large ($\sim 10,000$ pixel) submillimeter-wave
bolometer arrays presents a novel optical design problem---how can such
arrays be fed by diffraction-limited telescope optics where the primary
mirror is less than 100,000 wavelengths in diameter?  Standard
Cassegrain designs for radiotelescope optics exhibit focal surface
curvature so large that detectors cannot be placed more than 25 beam 
diameters from the central ray.  The problem is worse for Ritchey-Chretien 
designs, because these minimize coma while increasing field curvature.  
Classical aberrations, including coma, are usually dominated by
diffraction in submillimeter-wave single dish telescopes.  The telescope
designer must consider (1) diffraction, (2) aberration, (3) curvature of
field, (4) cross-polarization, (5) internal reflections,  (6) the effect 
of blockages, (7) means of beam chopping on- and off-source, 
(8) gravitational and thermal deformations of the primary mirror, 
(9) the physical mounting of large detector packages, and (10) the 
effect of gravity and (11) vibration on those detectors. 
Simultaneous optimization of these considerations in the case of large 
detector arrays leads to telescopes that differ considerably from 
standard radiotelescope designs.  Offset optics provide flexibility 
for mounting detectors, while eliminating blockage and internal reflections.  
Aberrations and cross-polarization can be the same as on-axis designs 
having the same diameter and focal length.  Trade-offs include the 
complication of primary mirror homology and an increase in overall cost.  
A dramatic increase in usable field of view can be achieved using shaped 
optics.  Solutions having one to six mirrors will be discussed, 
including a possible six-mirror design for the proposed South Pole 10m 
telescope.
\end{abstract}


  \section{INTRODUCTION} 

It will be possible in the coming decade to construct arrays of
submillimeter-wave detectors having $10^3$ to $10^4$ 
pixel elements\cite{mather98}.
This is an exciting prospect because there is important science to
be done at these wavelengths, observational work that requires deep imaging
capability over large areas of sky.  There is, however, a problem: an
array of $N \times N$ pixels requires a telescope whose field of view
is larger than $N$ beams in diameter, and no such submillimeter-wave
telescope exists for $N \gs 25$.  Design of such a telescope is a 
novel optical problem which differs significantly from the optical
design problems encountered for visual wavelength telescopes,
essentially because submillimeter-wave telescope apertures 
are only $\sim 10^5$ wavelengths across.

The problem can be summarized thus:
\begin{enumerate}
\item{An aperture  $D \sim 10 \, \mathrm{m}$ is needed to provide
the sensitivity and resolution to meet science goals.}
\item{If a detector array has $100 \times 100$ pixels, the
telescope field of view must be at least $\sqrt{2} \cdot 100 \cdot
\lambda/D \sim 0.8\deg $ in diameter.}
\item{Visual wavelength telescopes 10 m in diameter don't
have well corrected fields of view which are $0.8\deg$ in diameter
because of optical aberrations. Visual wavelength telescopes
can feed very large CCD arrays, but only because they have $D/\lambda > 10^6$.}
Many pixels then fit within a field of view that is $\sim 0.3\deg$ in
diameter.
\item{Radio and submillimeter telescopes 10 m in diameter with fields
of view $0.8\deg$ in diameter have 
large and mechanically awkward secondary and ancillary optics.}
\end{enumerate}

This paper discusses aspects of this problem and offers some
design solutions.  In \S2, the science goals are discussed, to
indicate the type of observation needed.
In \S3, submillimeter telescope designs are compared in
sensitivity and resolution, to show that design and construction 
of a wide field submillimeter telescope is desirable.
The design of wide-field submillimeter telescopes
having one or more optical elements is treated in \S4.

\section{SUBMM SCIENCE WITH LARGE DETECTOR ARRAYS} 

\begin{figure}[tbh!]
\begin{center} {
\leavevmode
\epsfxsize=4.00in
\epsfbox{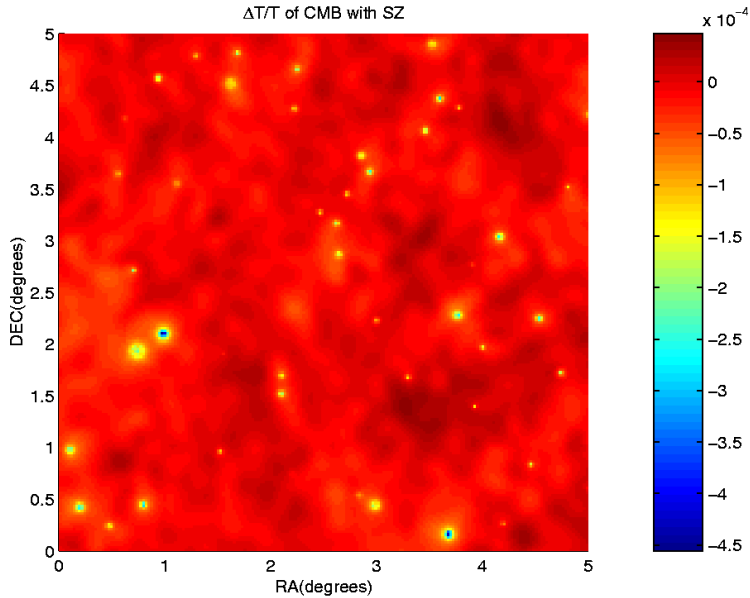}} \\
\parbox[t]{6.09in}{
\caption[Simulated Fine Angular Scale CMBR Image]
{{\bf Simulated Fine Angular Scale CMBR Image\cite{peterson99}}. Simulated CMBR structure on a twenty-five square degree region of sky.  The faint
extended structure is primary CMBR anisotropy, while the strong localized sources are
secondary anisotropy due to the Sunyaev-Zel'dovich effect in galaxy clusters.  In
this model, S-Z signal dominates at fine angular scales.  The simulation does not include
the S-Z filaments expected from the collapse of 100 Mpc structures.
\label{fig:goodmap} }
}
\end{center}
\end{figure}

\paragraph{Primary cosmic microwave background anisotropy at arcminute scales.}
Current CMBR observations\cite{whitewebpage} show a clear maximum in the anisotropy spectrum
at a spatial scale near 1\deg.
The damping tail region at arcminute scales provides a test
of the acoustic oscillation model.
The shape of the damping tail is a manifestation of the speed at which the
recombination process occurs, and the degree to which recombination is
mixed with re-ionization.
To study primary anisotropy at arcminute angular scales, it will be important
to simultaneously understand the secondary contribution to anisotropy due to
Sunyaev-Zel'dovich (S-Z) distortion in galaxy  clusters between the
recombination last scattering
surface and the observer.  
The
200-300 GHz range of frequencies are needed to allow spectral separation of
thermal S-Z effect from primary anisotropy.
Since there are 
$10 ^ 4$ to $10 ^ 5$ visible clusters of galaxies, a small beam size 
is needed at 200-300 GHz
to locate the cluster-free regions of sky that
will be observed for primary anisotropy studies
(see Figure~\ref{fig:goodmap}).

\begin{figure}[tbh!]
\begin{center} {
\epsfxsize=4in
\leavevmode
\epsfbox{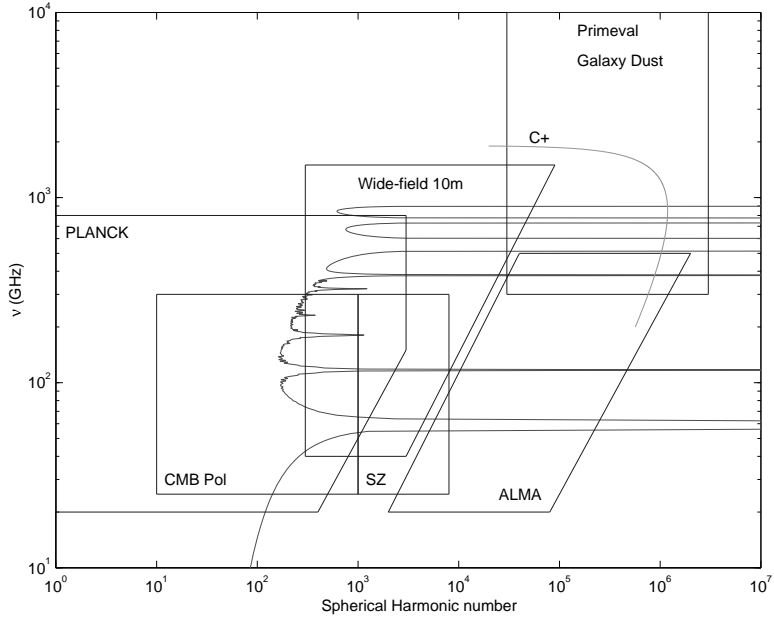} } \\
\parbox[t]{6.09in}{
\caption[A Comparison of Telescopes for CMBR]
{{\bf A Comparison of Telescopes for CMBR\cite{peterson99}}.  Shown in angular scale $l$ vs. observing frequency $\nu$ are science
targets: CMBR polarization, Sunyaev-Zel'dovich effect, dust emission by primeval galaxies, and
the fine structure line of ionized carbon from galaxies.  Also shown are the domains of
three instruments: the Planck spacecraft, a wide-field 10m, and the Atacama Large Millimeter
Array (ALMA).  The jagged curve is a comparison of atmospheric noise to noise in a
spacecraft environment.  On this curve, the South Pole winter atmospheric emission fluctuations
equal the photon fluctuations in a spacecraft environment (assumed to be caused by the brightness
of the sky plus photons emitted from a 1\% emissivity 70 K surface).  Observations above and to the
left of this curve derive a noise benefit from being in space.  Cluster S-Z observations fall to
the right of the noise boundary, and are best done with a wide-field 10m.  
The NASA Cosmic Microwave Background Future
Missions Working Group\cite{peterson99} recommends that 
most CMBR polarization observations be done from space, 
but high spatial frequency CMBR polarization can also be measured with the 10m
telescope.  
\label{fig:expplot} }
}
\end{center}
\end{figure}

\paragraph{Sunyaev-Zel'dovich effect in known clusters.}
As CMBR photons travel from the surface of last scattering
to the observer, secondary anisotropies can arise due to the
interaction of the CMBR photons with intervening matter.
Of particular interest is the S-Z effect, which
occurs when CMBR photons travel through a cluster of galaxies\cite{birkinshaw99}.
Approximately 10\% of the total mass of rich clusters of galaxies
is in the form of hot ($\sim 10^8 \, \rm K $) ionized plasma.
Compton scattering of CMBR photons by electrons in this intra-cluster
plasma can result in an optical depth as high as 0.02, resulting in
a distortion of the CMBR spectrum at the mK level.

The thermal component of the S-Z effect can be used in
combination with X-ray data to provide a measure of the Hubble
constant ($H_0$). In addition, when combined with a measurement of
electron temperature, the ratio of the kinematic
and thermal component amplitudes provides a direct measurement of the cluster's
peculiar velocity relative to the rest frame of the CMBR.
The observed surface brightness difference of both the thermal and kinematic
components is independent of the cluster redshift, as long as the cluster is
resolved.  Clusters are large objects, typically of order $1\,$Mpc, and
subtend an
arcminute or more at any redshift, so all clusters will be resolved with 
a 10m telescope.  
Accurate S-Z measurements can be made
throughout the Universe, all the way back to the epoch of formation of
the hot intra-cluster
gas.
 
\paragraph{Sunyaev-Zel'dovich effect blank-sky survey.}

The fast survey capability of a wide-field submillimeter telescope
 will allow detection of
S-Z effect in blank-sky searches.
Detections of small-scale decrements in CMBR intensity not
associated with
known clusters may have been discovered by Ryle Telescope\cite{jones97} and VLA
\cite{Richards97} observations.  These may be due to the S-Z effect in very
distant clusters.  As shown in Figure~\ref{fig:goodmap}, serendipitous detection 
of distant clusters will necessarily occur in studies of arcminute-scale CMBR
primary anisotropy.
These detections,
both with and without additional X-ray data\cite{holder99}, allow
powerful tests of cosmological models and galaxy formation
theories.  In particular, the cluster counts are
sensitive to $\Omega_\Lambda$ and provide an independent check
of type Ia supernova results.  

S-Z effect also results from low density, warm baryonic gas
between clusters\cite{cen99}.  Models of structure formation 
predict that most of the baryonic matter in
the Universe is located in intra-cluster filaments which are
responsible for the Ly-$\alpha$ forest seen in quasar spectra.
The S-Z effect caused by the filaments can be directly imaged, 
permitting study of the filaments
in the spaces between quasars.

\paragraph{Continuum detection of high redshift protogalaxies not detectable in the 
visible or near-IR.} Most of 
the luminosity resulting from the collapse energy of galaxies and 
the first generations of stars may not appear at visual or 
near-infrared wavelengths (even in the rest frame of the distant 
galaxy), but may instead be reradiated by dust\cite{meurer98,pearson96}.
Detection of the cosmic far-infrared background 
radiation\cite{puget96} (CFIRBR)
is evidence that most of the energy released in the
initial collapse of galaxies and the creation of metals appears 
at the current epoch as submillimeter-wave radiation.
In a recent review of
work on the Hubble Deep Field (HDF),  P. Madau wrote that ``the
poorly constrained amount of starlight that was absorbed by dust and reradiated
in the far-IR at early epochs represents one of the biggest uncertainties in our 
understanding of the evolution of luminous matter in the universe"
\cite{madau98}.  

Within the next decade,
submillimeter studies of known sources at high redshift will be
carried out with powerful interferometer arrays, the SMA and the ALMA, at least
at longer submillimeter wavelengths.
A fundamental contribution of wide-field single dish submm telescopes
will be to detect and locate 
high redshift sources by studying the structure
of the CFIRBR 
on a scale of $\sim 10''$.
Such work has already begun on Mauna Kea.
At 850\micron wavelength, SCUBA\cite{smail97,hughes98} has observed
a 10 square arcminute field to an rms noise
level of 2 mJy beam$^{-1}$ and 
a 6 square arcminute field centered on the HDF to an rms noise level of 0.5 mJy beam$^{-1}$.
These surveys show detected source number densities of about 1 per square
arcminute, where at least some sources are at $z \sim 0.3$ to $4$ and have inconspicuous (or possibly
no) optical counterparts\cite{smail99,richards99a}.  
Some models of protogalaxies consistent with a protogalactic origin for the CFIRBR\cite{guiderdoni98} 
predict hundreds of sources per square degree which are detectable at
the $\ge 1$ mJy level in the submillimeter, but which have optical counterparts 
below the detection limit of the HST HDF image.

\section{TELESCOPE SENSITIVITY}

\begin{table}[tbh!]
\begin{center}
\caption[Continuum Sensitivity of Submillimeter Telescopes] 
{{\bf Continuum Sensitivity of Submillimeter Telescopes\ \ } 
\\
\label{table:continuum} }
\begin{tabular}{lrrrrrrrrr}
Telescope & {$A^{\rm a}$} & {$R^{\rm b}$}& 
            {$S^{\rm c}$}& \multicolumn{3}{c}{NEFD$^{\rm d}$} 
            & \multicolumn{3}{c}{Time in hours to survey} \\
& $({\rm m^2})$ & $( '')$ &  & \multicolumn{3}{c}{$({\rm mJy \, s^{1/2}})$} & 
             \multicolumn{3}{c}{1 square degree at 1 mJy} \\
& & &  & $850 \mu \rm m$ & $450 \mu \rm m$ & $350 \mu \rm m$ & $850 \mu \rm m$ & $450 \mu \rm m$ & $350 \mu \rm m$ \\
\\
wide-field 10 m & 79 & 11 & 200 & {\it 60} & {\it 64} & {\it 74} & {\it 18} & {\it 20} & {\it 27}\\
wide-field 30 m & 711 & 4  & 22 & {\it 7} & {\it 7} & {\it 8} & {\it 2} & {\it 2} & {\it 3}\\
AST/RO & 2 & 65 & 92 & {\it 2160} & {\it 2300} & {\it 2660} &  ${\it 5.1 \times 10^4}$ & ${\it 5.8 \times 10^4}$ & ${\it 7.7 \times 10^4}$\\
JCMT & 177 & 7 & 5 & 80 & 700 & {\it 760} & $1.3 \times 10^3$ & $9.8 \times 10^4$& ${\it 1.2 \times 10^5}$\\
CSO & 79 & 11 & 11 & {\it 150} & 2000 & {\it 2200} & {$\it 2.0 \times 10^3$} & $3.6 \times 10^5$ & ${\it 4.4 \times 10^5}$\\
SOFIA & 5 & 44 & 50 &  & {\it 200} & {\it 200} &  & {\it 800} & {\it 800}\\
FIRST & 7 & 32 & 4.3 &  & {\it 54} & {\it 54} &  & {\it 678} & {\it 583}\\
``submm CBI"$^{\rm e}$ & 8.3 & 11 & 4 & {\it 297} & {\it 2529} & {\it 2768} & ${\it 6.2 \times 10^3}$ & ${\it 1.6 \times 10^6}$ & ${\it 1.1 \times 10^7}$ \\
SMA & 227 & 2 & 0.14 & {\it 134} & {\it 1142} & {\it 1250} & ${\it 3.6 \times 10^4}$ & ${\it 9.3 \times 10^6}$ & ${\it 1.9 \times 10^7}$ \\
MK array$^{\rm f}$ & 483 & 0.5 & 0.02 & {\it 59} & {\it 719} & {\it 790} & ${\it 4.9 \times 10^4}$ & ${\it 2.6 \times 10^7}$ & ${\it 5.2 \times 10^7}$ \\
MMA & 2010 & 0.2 & 0.08 & {\it 7} & {\it 45} & {\it 52} & ${\it 230}$ & ${\it 3.4 \times 10^4}$ & ${\it 7.5 \times 10^4}$\\
ALMA & 7000 & 0.2 & 0.06 & {\it 2} & {\it 12} & {\it 15} & ${\it 19}$ & ${\it 2.4 \times 10^3}$ & ${\it 6.4 \times 10^3}$

\end{tabular}
\end{center}
\noindent 
{
Notes: 
\hfill\break
{a.}\,{Telescope area ($\rm m^2$)} \hfill
\break
{b.}\,{Resolution element ($\rm arcsec$) for $\lambda = 450 \micron$.
Resolution element scales as $\lambda$.} \hfill
\break
{c.}\,{Instantaneous sky coverage ($\rm arcmin^2$) for $\lambda = 450 \micron$.
Instantaneous Sky Coverage 
scales as $\lambda^2$ for interferometers,
is independent of $\lambda$ for most single-dish instruments, and is 4.3,
5.0 and 2.7 arcmin$^2$ at $480\micron$, $350\micron$ and $250\micron$,
respectively,
for FIRST (G. Pilbratt, personal communication).  }
\hfill
\break
{d.}\,{Noise Equivalent Flux Density, the sensitivity to point sources whose
positions are known.
Numbers in italics are predicted sensitivities; numbers not in italics are
measured, on-the-telescope values and are subject to downward revision with
improved techniques.  Predicted sensitivities are optimistic in the
sense that in all cases they are near the thermal background limit, a
limit that has not yet been achieved in practical
submillimeter-wave bolometer systems.
This table is based on the work of Hughes and Dunlop\cite{hughes97a}.}
\hfill
\break
{e.}\,{A hypothetical submillimeter-wave array with the configuration of the CBI; the actual 
CBI operates at $\lambda \sim$ 1 cm.}
\hfill
\break
{f.}\,{Mauna Kea array consisting of the SMA, the CSO, and the JCMT}
}
\end{table}

\begin{figure}[tbh]
\begin{center} {
\leavevmode
\epsfxsize=7.0in
\epsfbox{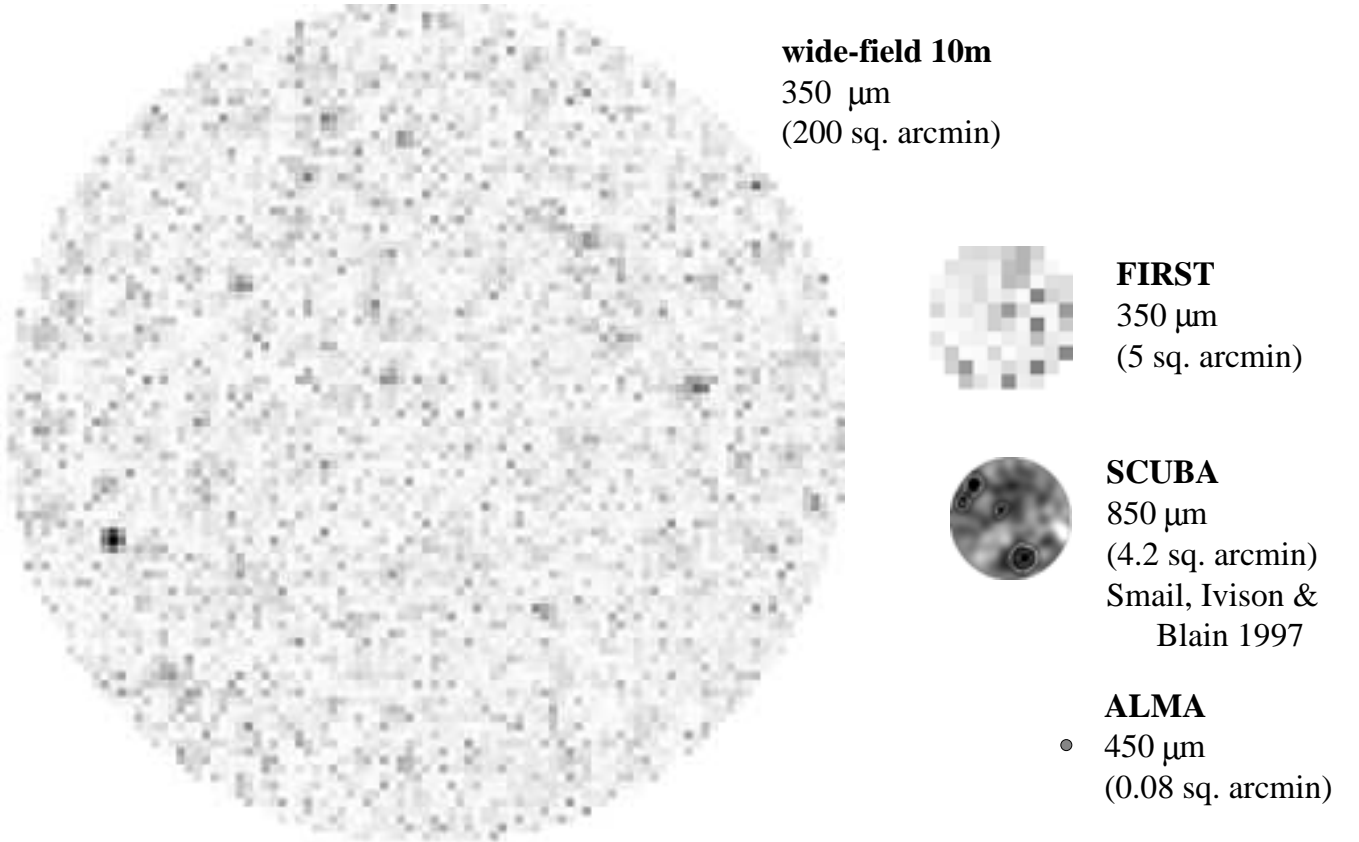} } \\
\parbox[t]{6.09in}{
\caption[Image Sizes of Submillimeter-wave Telescopes]
{
{\bf Image Sizes of Submillimeter-wave Telescopes.\ \ }
Simulated images from a wide-field single dish 10m and 
FIRST are compared to SCUBA results
\cite{smail97} and the image size of the ALMA. The simulated data for
the single dish and FIRST are the Guiderdoni et al.
protogalaxy Scenario B model\cite{guiderdoni98} 
observed to an
rms noise level of 1 mJy per pixel with a broad-band bolometer array.
This noise level is attainable with about two hours of on-source time
at the wide field single dish 
and 20 minutes of on-source time with FIRST.
It is assumed that the protogalaxies are Gaussian distributed (i.e., not clustered).
A black pixel corresponds to 78 mJy, a white pixel is 0 mJy, and the
greyscale is linear.  The single dish plot shows over 500 individual sources
at the $5 \sigma$ level.
\label{fig:simul} }
}
\end{center}
\end{figure}

Table~\ref{table:continuum} shows the continuum
sensitivity and beam size of submillimeter-wave telescopes and
illustrates the strength of wide field single dish for deep continuum surveys.
The NEFD values in this table for each of the instruments originate if possible from 
the scientific group operating or proposing that instrument; they are a
mixed bag of actual results on the telescope and calculations, some 
of which are based on optimistic assumptions.
A surprising implication of Table~\ref{table:continuum} is that a wide
field single dish
is faster at submillimeter detection of protogalaxies than the
ALMA and FIRST.  Is this plausible, given that these instruments are at least
an order of magnitude more expensive?   Suppose we are conducting
a large-scale survey at $\lambda450 \micron \ $  for point-source objects at an rms 
flux density $F_{\mathit limit}$.  The time required for $1 \sigma$ detection 
in one pixel of a map of the sky is given by the radiometer equation
\begin{equation}
\label{eq:radiometer}
t_{\mathit limit} = {{(2kT_{\mathit sys})^2}\over{B (\eta A F_{\mathit limit})^2 }} = 
\left( {{\mathrm NEFD }\over{F_{\mathit limit}}} \right) ^2 ,
\end{equation}
where
$$
T_{\mathit sys} = \left( T_{\mathit receiver} + \eta T_{\mathit atmosphere} [1 - {\rm e}^{-\tau
}] + [1 - \eta] T_{\mathit ambient} \right)
{{{\rm e}^{\tau }}\over { \eta}}
$$
is the atmosphere-corrected effective system temperature, 
$A$ is the total collecting area of the telescope,
$\eta$ is telescope efficiency, 
$\tau$ is the atmospheric opacity at the elevation
angle of the observation, and $B$ is the pre-detection
bandwidth.  $T_{\mathit atmosphere}\approx T_{\mathit ambient}\approx 200\rm \, K$ at
the South Pole and $\approx 260 \, \rm K$ at mid-latitude sites.
Here we neglect polarization, relative sideband gain,
the distinction between telescope efficiency and aperture efficiency,
digitization and data processing corrections:
these introduce factors of order unity which have been considered in
Table~\ref{table:continuum} but which can be neglected in the present plausibility argument.
We assume that the radiometer is switched rapidly enough to
filter out sky noise.
Equation~\ref{eq:radiometer} is
then approximately correct regardless of whether the system is
heterodyne or bolometric, or whether the collecting area, $A$, is arranged
in multiple antennas or a single dish.  
All telescopes will be designed for reasonably high efficiency, $\eta \sim 0.9$.
Future submillimeter-wave systems will be
background-limited, meaning that the $T_{\mathit receiver}$ term will be
smaller than the atmospheric ($T_{\mathit atmosphere}$) and telescope
background ($T_{\mathit ambient}$) terms.  
When $\tau > 1$, as it often is for ground-based submillimeter-wave observations,
the opacity-correction term, $e^ \tau$, dominates all other effects: 
$
T_{\mathit sys} \approx {\mathrm e}^\tau  ( T_{\mathit atmosphere} + T_{\mathit receiver}/\eta ),
$
and $t_{\mathit limit} \propto {\mathrm exp}(2 \tau)$; 
in this case relatively small improvements in $\tau$
make a large difference.  
At $\lambda 450 \micron $ and $45 \deg$
elevation 
the 25\% opacity values in winter
are
$\tau \ls 0.36$ at the South Pole and
$\tau \ls 0.79$ at Chajnantor.
When $\tau$ is small or zero, for example in space or on the ground 
at centimeter wavelengths, then the 
telescope background $(1 - \eta) T_{\mathit ambient} $ dominates. 
For a single dish located at the 
South Pole, $T_{\mathit sys}\simeq 1000\,\rm K$ (a value that is often surpassed
on AST/RO),
for the ALMA, $T_{\mathit sys}\simeq 2500\,\rm K$ (this high system temperature is
due mostly to atmospheric opacity),
and for FIRST, $T_{\mathit sys}\simeq 50\,\rm K $ (the low
background of a cooled telescope in space).
For the wide-field single dish
$B\simeq 100\ {\rm GHz}$, 
for FIRST, 
$B\simeq 200\ {\rm GHz}$, 
whereas for
the ALMA $B\simeq 16 \ {\rm GHz}$.  We therefore conclude that
the time required to 
achieve a $F_{\mathit limit} = 1 {\rm mJy}$ noise level, $t_{\rm mJy}$, is about four hours for the
wide-field single dish, forty minutes for FIRST, and one minute for the ALMA.  
To reach an rms
noise of, say, 0.1 mJy would require 100 times longer 
integration in each case.   ALMA and FIRST are both 
faster than the ground-based single dish at detecting individual sources.

Now we note that the area of sky covered
by the field of view
during this integration time, the ``instantaneous sky coverage"
($S$ in Table~\ref{table:continuum}, see Figure~\ref{fig:simul}), is
very different for the three telescopes.  
For the wide field 10m, $S$ is over an order-of-magnitude larger than FIRST and nearly 
four orders-of-magnitude larger than the ALMA (see Figure~\ref{fig:simul}).
An efficient configuration of $n$ diffraction-limited detectors (bolometer pixels
or heterodyne receivers) on a telescope of total area $A$ will yield an
instantaneous sky coverage of
$
S \approx {{ n \lambda^2}\over{ \eta A}}.
$
The speed at which the 
the sky can be mapped is $S/{t_{\mathit limit}}$.
A figure of merit for blank sky surveys can therefore be defined:
\begin{equation}
\label{eq:figureofmerit}
{\mathrm survey \  speed} 
\propto {\mathrm figure \ of \ merit} \equiv {{n B \eta A  } \over {T_{\mathit sys}^2}}.
\end{equation}

\begin{table} [bt!]
\begin{center} {
\parbox{6.3in}{
\caption[Bolometer Array Instruments] 
{{\bf Bolometer Array Instruments\ } 
\\
\label{table:bolometers} }
\begin{tabbing}
Instrument \qquad \= in development \qquad \= Number of \qquad \=  http://pioneer.gsfc.nasa.gov/public/safire   \kill
Instrument \> Current \> Number of  \> Web Reference \\[0pt]
\> Status  \> Detectors, $n$  \\[3pt]
\rule[6pt]{6.3in}{0.2pt}\\[0pt]
Instrument \qquad \= in development \qquad \= Number o\=f \qquad \=  http://pioneer.gsfc.nasa.gov/public/safire   \kill
SCUBA\> operational \>\>      128 \' \> {\tt http://www.jach.hawaii.edu/JACpublic/JCMT/scuba } \\[2pt]
BOLOCAM\> in development \>\>      151 \' \> {\tt http://binizaa.inaoep.mx/pub/ins/pres\_cam} \\[2pt]
SAFIRE\> in development \>\>    2,048 \' \> {\tt http://pioneer.gsfc.nasa.gov/public/safire } \\[2pt]
SPECS\> proposed        \>\>   60,000 \' \> {\tt http://www.gsfc.nasa.gov/astro/specs} \\[2pt]
\rule[5pt]{6.3in}{0.3pt} \\[0pt]
\end{tabbing}
} 
} \end{center}
\end{table}

The trade-offs in survey speed between a wide-field single-dish
telescope and an array instrument like the ALMA
can be estimated by evaluating Equation~\ref{eq:figureofmerit}.  Suppose
the single dish telescope has a focal plane array containing $N \times N$ detectors,
and the interferometer consists of $E$ antenna elements, each with the same diameter
as the single-dish telescope, and each having one heterodyne receiver.  
Then
$
S_{\mathrm dish}= N^2 S_{\mathrm interferometer},
$
$
A_{\mathrm interferometer} = E A_{\mathrm dish},
$
$
n_{\mathrm interferometer} = E ,
$
and
$
n_{\mathrm dish} = N^2 .
$
If the two instruments have the same $T_{\mathit sys}$, the single dish
telescope will be faster in survey 
\enlargethispage*{1000pt}
mode by a factor 
$$
{{N^2}\over{E^2}}{{B_{\mathrm dish}}\over{B_{\mathrm interferometer}}} .
$$
\pagebreak

Comparing the wide-field 10m with the ALMA, $N\simeq 100$, $E\simeq 80$, and
${{B_{\mathrm dish}}/{B_{\mathrm interferometer}}}\simeq 8$ .
{\em 
The single-dish 10m
is at least an order-of-magnitude faster than the ALMA at submillimeter-wave
sky surveys.}
It might be argued that $N^2$ detectors represents a lot of complex
electronics, and that for large numbers it may be easier to 
build $E$
whole antennas than $N^2$ detectors---however, any interferometer
will necessarily have $E^2 $ correlators; these correlators
are likely to be more complex than the photolithographically-produced 
detector plus amplifier plus multiplex needed for each
of the $N^2$ pixels of a bolometer array.
Table~\ref{table:bolometers} shows the number of detectors, $n$, in some
current and proposed millimeter and submillimeter instruments.  The
large jump in $n$ between the current version of BOLOCAM and SAFIRE reflects
the development of cryogenic multiplex readouts.
Compared to single-dish maps,
maps made by the interferometric instrument have much higher resolution 
and positional accuracy, but the rate at which 
the maps are made is at least an order-of-magnitude slower and the initial
capital cost is at least an order-of-magnitude higher.  
It is a waste of scarce resources to map large areas of blank sky
with the ALMA.  
Note too that at a survey speed of $\sim 2.4 \times 10 ^3$ hours per square
degree, the ALMA will not be able to survey more than a hundred square degrees
during its operating lifetime, but there are many thousands of square
degrees containing potentially interesting sources.
The best strategy to detect and study protogalaxies is to survey the sky
quickly with the relatively inexpensive single dish and then study the
detected sources in detail with the ALMA.

For speed and coverage in deep surveys, 
a wide-field telescope can pursue an
optimization strategy:
\begin{enumerate}
\item  Build a reflector whose beamsize is equal
to the spatial scale of interest. 
\item Make the telescope losses, blockage and spillover as small as possible
at the frequencies of interest.
\item Make the field of view as large as possible. 
\item Populate that field of view with as many broadband detectors
as possible.
\item Place the telescope at the best possible site.
\end{enumerate}
Such a wide-field single dish is an ``optimal design"
for the discovery of objects at spatial frequencies near 
$\frac{1}{2} D / \lambda$.

\section{Telescope Design}

The physical size of a single pixel in an array of bolometer detectors
must be half a wavelength or more in order to efficiently couple
power into the detector.
If the pixel size is as small as a wavelength, the beam impinging on the
bolometer array must be highly convergent (have an $f/d \equiv
F_{\mathit{detector}} \ls 1$),
since the beamwaist diameter\cite{goldsmith98} at the detector is $\approx \lambda F_{\mathit{detector}}/\pi$.
A submillimeter-wave detector array 
containing $N^2$ pixels will necessarily be $N F_{\mathit{detector}}/\pi$ 
wavelengths
across, or about 3 cm for a $100 \times 100$ array.  
Suppose the telescope which will feed the array has
an aperture $D$ and an effective focal length $f_e$.
Then a diffraction-limited beam on the sky will have an angular width
of $\lambda/D$.  The telescope field of view must be about
$2 N$ beams across to fully feed the array and
allow for beam chopping.  Let the desired field of view
be $\zeta N \lambda/D$ where $\zeta \sim 1.5$ is an arbitrary parameter
to allow for illumination of the corners of the array and the
extra field of view needed for chopping.
At the surface of the detector, this field
of view has a physical size
$ \zeta N \lambda f_e /D = \zeta N \lambda F_e$.  The telescope must provide a
field of view at least this large such that aberrations
are small compared to the pixel size.

\subsection{Prime focus} A 
prime focus feed is the simplest possible design.  
The detector array is designed
for a focal ratio
$F_{\mathit{detector}}$, so perhaps it can be matched 
by simply placing it at the focus of a single reflector having a focal ratio
$F \equiv f/D = F_{\mathit{detector}}$.  This will, in fact,
match the central pixel and provide the correct plate scale
(overall magnification of the telescope in, e.g., millimeters per
arcsecond)
to match the immediately surrounding pixels to adjacent beams
on the sky.  

A field of view the size of the array must have sufficiently
small aberrations.
The dominant aberrations for a single paraboloidal reflector are
coma and astigmatism.
The angular coma is
$
\theta_c = {{\theta}\over{16 F^2}} ,
$
and the angular astigmatism is
$
\theta_a = {{\theta^2}\over{2 F}} ,
$
where $\theta_c$ and $\theta_a$ are the angular blur due to coma 
and astigmatism for a beam at angle
$\theta$ from the center of the field\cite{schroeder87}.
The angular blur due to diffraction is 
$
\theta_d = {{\lambda}/{D}},
$
so the coma blur is less than the diffraction blur for
field angles
$
\theta \ls {{16 F^2 \lambda }/{D}},
$
and the astigmatic blur is less than the diffraction
blur for field angles
$
\theta \ls \sqrt{{{2 F \lambda }/{D}}}.
$
These aberrations place two constraints on the
mirror optics:
\begin{equation}
\label{eq:comasingle}
F \gs \sqrt{{\zeta N}\over{32}},
\end{equation}
and 
\begin{equation}
\label{eq:astigmatismsingle}
f = FD  \gs {\frac{1}{2}} \, \zeta^2 N^2 \lambda.
\end{equation}

For a submillimeter-wave telescope operating with a large
detector array, Equation \ref{eq:comasingle} requires that
$F \gs 2.2$, while 
Equation \ref{eq:astigmatismsingle} requires that
$f \gs 10{\rm m}$.  
A reflector satisfying these constraints will have an
acceptable image, but it
makes for a telescope
design with five disadvantages:
\begin{enumerate}
\item{The $F$ number is much bigger than the
values usually chosen for radiotelescopes, resulting
in a large, elongated structure.  Most
radiotelescopes have $F \sim 0.5$ for the primary mirror.}
\item{Since there is no chopper mirror, there is no way
to rapidly switch the position of the array on the sky---the
entire telescope must be moved.
}
\item{The detector package must tilt with the movement of
the telescope.  If the telescope system is ground-based, the
direction of gravity
will rotate by $90\deg$ with respect to the detector and its cryogenics
system as the telescope moves in elevation.
}
\item{The detector package blocks the primary mirror.}
\item{Since $F_{\mathit{detector}} = F > 2.2$, the size of the detector package 
and the size of the opening into the cryogenics will
be larger than optimal.}
\end{enumerate}

Problem 4, the blockage of the primary, can be avoided by using
an off-axis section of the parabola.  This brings with it additional
problems of cross-polarization and asymmetric illumination, but
for a system with large $F$ number these will not be large effects.
An off-axis single reflector has about the same aberrations as an
on-axis reflector with the same focal length and twice the
aperture, so the $F$ number of the off-axis system must be about
twice as large as the value given by Equation \ref{eq:comasingle}.

Problems 2 and 3 could be obviated by placing the telescope in space.
Imagine a spacecraft with a large submillimeter detector package
fed by a $\sim 2$ m diameter off-axis mirror 
with a $\sim 10$ m focal length.
This instrument would scan in a ``stare" mode
and would produce maps with a resolution about $1'$.
 
A possible variant of the single mirror offset telescope is
a primary mirror with a flat or nearly flat secondary.  This
mirror could fold the beam back below the primary mirror
and act as a chopper---a kind of folded Newtonian.  Such
a mirror might also act as an aspheric corrector, providing
a better compromise between spherical aberration and coma,
or providing correction for a spherical primary. 

\subsection{Cassegrain and Gregorian designs} 
The addition of a second curved mirror brings
considerable additional freedom and improved performance to telescope design.
Cassegrain designs, with
a concave primary and a convex secondary,
are ubiquitously successful from visual
to radio wavelengths.
The reasons for this success differ at different wavelengths.
At radio wavelengths, the design criterion is to provide
high aperture efficiency and minimal blockage for a single beam
or a few beams near the center of the field of view.
At visual wavelengths, the design criterion is to provide
a field of view about $10'$ across which is free of coma, the
otherwise dominant aberration---this is accomplished by the
aplanatic Cassegrain (or
Ritchey-Chretien) which reduces coma to less than a
second of arc at the expense of increased image curvature.
The submm telescope problem is novel because a large field of view
is needed in a situation where diffraction
dominates the classical
aberrations, and it is curvature of field rather than
coma which tends to limit the field of view.
These issues were discussed at an earlier SPIE conference\cite{stark98b}.

A telescope having only two 
mirrors 
would not be used to feed a submillimeter-wave detector array directly.
Large submillimeter-wave detector arrays require 
a small value for $F_{\mathit{detector}}$ 
($\ls 3$) so that the detector size is
manageable, whereas  two mirror telescopes
have large values of $F_e$ ( $\approx { 1.5 \cdot \mathrm{ (focal\ length\ 
of\ primary)/(diameter\  of\ secondary)}} \sim 30$).
At least a few additional optical elements are
needed as focal reducers.  If the primary and secondary produce a
sufficiently corrected image with a
suitably large field of view, additional optics can feed that image
onto the array.

Let $R_1$ and $R_2$ be 
radius of curvature at the vertex of
the primary and secondary mirrors, 
respectively.
The focal length of the primary mirror is
half the radius of curvature, $f_1 = R_1/2$.
Let $m \equiv 2 f_e/R_1$ be
the magnification of the secondary.  
If the secondary is made as small as it can be without
vignetting the center of the field, then the ratio of the diameter of
the secondary, $D_2$, to the diameter of the primary, $D_1$, is
$
k \equiv {{D_2}\over{D_1}} = {{R_2}\over{R_1}} \, {{m-1}\over{m}}.
$
The distance from the back of the primary to the focus of the
two-mirror system is
$
f_1 \beta \equiv f_1 [k (m+1) -1] \, ,
$
where $\beta$ is the ``scaleless backfocal distance" and is
either positive or negative depending on whether the focus
is behind or before the primary mirror.
These variables and their sign conventions are
as defined by Schroeder\cite{schroeder87}.

Consider first a conventional on-axis design. 
The beam passes through a hole in the primary mirror,
and the hole in the primary should be no larger than
$D_2$, the diameter of the secondary.  The diameter of
the image is
\begin{equation}
\label{eq:Dvariable}
D_i \equiv \zeta N {{\lambda}\over{D_1}} f_e = \zeta N \lambda m F_1 .
\end{equation}
To avoid vignetting, $D_i$ must be smaller than or equal to the
diameter of the central
hole in the primary, which must be smaller than or equal to the diameter
of the secondary, $D_2$.  But as $D_2$ becomes smaller, $| m |$ becomes
larger and $D_i$ becomes larger.  The best that can be done is
to make $D_2$, $D_i$, and the hole in the primary the same size.
Letting $D_2 = D_i$ in Equation \ref{eq:Dvariable} and substituting
for $k$ and $m$ yields a quadratic equation that can be solved for
$D_2$:
\begin{equation}
\label{eq:D2solution}
D_2 \geq D_i = {\frac{1}{2}} \zeta N \lambda F_1 \left [
{\sqrt{{{4(1+\beta)D_1}\over{\zeta N \lambda F_1}}+1\;}}\; - \; 1 \right ]
\approx 
{\sqrt{(1+\beta)\zeta N \lambda f_1 }} \ .
\end{equation}
This is a minimum size for the secondary mirror 
of a two mirror telescope to feed an $N \times N$ array at
wavelength $\lambda$.  For a 10m diameter $F_1 = 0.4$ 
telescope feeding an $N=100$ array at $\lambda = 1 \mathrm{mm}$, 
$D_2 \gs 1.1 \mathrm{m}$.  This is considerably larger than the
secondary mirrors usually fitted to Cassegrain systems.
In order to feed a $100 \times 100$ array, the CSO and
JCMT would have to be refitted with secondary mirrors (and primary
mirror openings) which are about twice as large as the current ones.
Antennas designed as elements of interferometric arrays
are particularly bad, since
they are highly optimized to feed a single beam with high
efficiency and therefore have minimal
secondaries.

\paragraph{Ancillary optics.}
If the value of $D_2$ satisfies Equation \ref{eq:D2solution},
it is possible to achieve a $\approx 1\deg$ diameter field of view
at $\lambda = 1 \, \mathrm{mm}$
with a $D_1 = 10 \, \mathrm{m}$ Cassegrain or Gregorian.  The
dominant aberration is curvature of field; such a telescope
has a surface of best images which resembles a 3 steradian cap
on a 1.5 m diameter sphere (concave toward the secondary in
a Cassegrain and convex toward the secondary in a Gregorian).

Additional optics are needed to match
a Cassegrain or Gregorian focus to a detector array, and these
can be used to correct some of the deficiencies in the quality
of focus.
In particular a field-flattening mirror is a mirror placed
at the telescope focus which has the same direction of
curvature as the focal plane.  
The surfaces of best images will be flattened
when the beam is refocused by subsequent optics, 

The ancillary optics in a wide field telescope
of conventional telescope design will be large---one to three
meters in diameter, in order to handle an image as large 
as that given by Equation \ref{eq:D2solution}, and the
further behind the primary backup structure they are
located, the larger they must be (since $D_i$ increases
with increasing $\beta$).  This creates problems in structural
design.  Large cavities with big mirrors need to be located 
within the support structure of primary mirror.
If the telescope has a Nasmyth focus,
the hole in the elevation bearing must be large.

\begin{figure}[tbh!]
\begin{center} 
\leavevmode
\epsfxsize=4in
\epsfbox{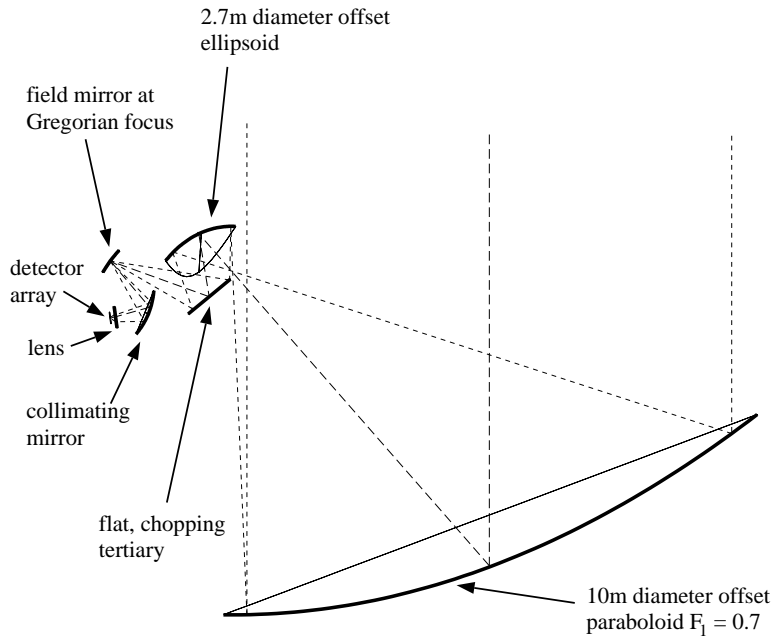}  \\
\parbox[t]{6.09in}{
\caption[Scale drawing of a wide-field submm telescope.]
{{\bf Scale drawing of a wide-field submm telescope.\ \ }
This is an offset Gregorian design with ancillary optics
which provide beam chopping and a focal reducer to
feed a large detector array.  All optical elements are
symmetric about the plane of the page; in a practical system
the fourth and fifth mirrors would fold the beam so that
it was parallel with the telescope's elevation axis at the point where
the beam entered the cryogenic dewar---this would allow the
dewar to rotate around the beam and not tilt with
telescope
elevation.  The primary and secondary are a 
conventional offset Gregorian design having
$D_1 = 10\, \mathrm{m}$, $F_1 = 0.7$ and
$\beta = -0.7$, except that the secondary is tilted to
satisfy the
Dragone\cite{dragone82} relation with
$i_0 = 0$.  The flat tertiary is at an image of the aperture.
The fourth and fifth mirrors are highly aspheric, offset, polynomial
surfaces calculated to correct aberrations at the detector array.
The fourth mirror is located at the Gregorian focus and acts as
a field lens to image the exit pupil of the telescope onto 
the aperture stop of a focal reducer consisting of
a collimating mirror and a lens.  This telescope
has a usable field of view about $1\deg$ at $\lambda = 1 \,
\mathrm{mm}$.
\label{fig:optics} }
}
\end{center}
\end{figure}

\paragraph{Offset optics.}
An off-axis design for the primary mirror helps solve this problem.
If the receiver does not block the primary, then $\beta$ can be
negative, placing the focus of the secondary in front of but below
the primary.  This reduces the overall focal length, $f_e$ and
thereby reduces the size of the image.
The ancillary optics are then both smaller and removed from the region
behind the primary mirror.  Figure \ref{fig:optics} shows an off-axis
Gregorian design that provides for beam chopping and has a $1\deg$
usable field of view.  This design is very nearly a $1/10$ model of
the Green Bank Telescope\cite{greenbanktelescope}, 
because it solves similar design problems
at approximately $1/10$ the wavelength.

There is no optical penalty for using offset optics.
Dragone\cite{dragone82} has shown that if the offset angles in an offset 
two mirror telescope are chosen correctly, then aberrations and 
cross-polarization effects in that offset 
telescope are the same as those in a conventional on-axis antenna with 
the same diameter
and focal length.  The beam efficiency, aperture efficiency, and 
sidelobe levels in the off-axis antenna
are better than those in the on-axis antenna, because in the on-axis design there
will be diffraction, reflection, and blockage from the secondary mirror 
and its supports.
An off-axis two mirror telescope with correctly chosen offset angles
will always be optically superior to a similar on-axis configuration. 

Avoidance of internal reflections is beneficial to the design of 
submillimeter-wave telescopes.  Any receiver or detector placed at the 
focus of the instrument will necessarily emit into the telescope some 
amount of submillimeter-wave power in the band of interest.  If there is a 
reflection in the system, for example at the secondary mirror some 6 
meters distant,
then a resonant cavity is formed whose modes are spaced at 
roughly 25 MHz intervals.
Also critical to submillimeter-wave telescope design is minimization of
variations in antenna thermal emissions as a function of chopper angle.
Chopper offsets can more easily be minimized in an off-axies design.
In an off-axis Gregorian the placement of the chopper at an image
of the primary allows the beam to be steered on the sky without
changing the illumination of the primary.

It is more expensive to build a telescope off-axis. Compared to an
on-axis design of the same aperture, there are twice as many types of
surface panels for the primary mirror.  The primary backup structure is
less symmetric, making the structural design for homology more complex
and increasing the number of dissimilar parts. These detriments are
mitigated by the modern use of computers in design and manufacturing.
A detailed consideration of the design problems for on-axis systems
will likely show that for a given cost, an offset design can provide
more pixels with a greater effective collecting area than an on-axis
design.

\section{CONCLUSION}

The salient results of this investigation are:
\begin{enumerate}
\item{There is important science to be done which requires
large-scale imaging at submillimeter wavelengths.}
\item{Single-dish telescopes with large detector arrays are
faster and more cost effective
at mapping large areas of sky than are other telescope types.
Such instruments are highly complementary to interferometric arrays,
detecting sources to be studied in detail by
interferometeric techniques.}
\item{None of the existing designs for submillimeter-wave telescopes
are capable of feeding large detector arrays.  In particular, the
secondary mirror must be large.}
\item{It is nevertheless possible to design a telescope which
will accommodate large detector arrays while providing for beam chopping,
if the design starts with a clean slate.}
\end{enumerate}

\acknowledgments     

I thank J. B. Peterson and C. Cantalupo 
of Carnegie-Mellon U. for 
Figures 1 and 2.
This work was supported in part by the Smithsonian Institution and
in part by the National Science Foundation
under a cooperative agreement with the Center for Astrophysical Research
in Antarctica (CARA), grant number NSF DPP 89-20223.  CARA is a National
Science Foundation Science and Technology Center.


\bibliographystyle{spiebib}
\bibliography{mybib,cara,astro98,stark}
 
  \end{document}